\documentclass[aps,pra,floatfix,twocolumn, superscriptaddress,10pt]{revtex4-1}

\usepackage{graphicx}
\usepackage{amsmath}
\usepackage{mathtools}
\usepackage{amssymb,bm}

\usepackage[svgnames]{xcolor}

\usepackage{hyperref}
\hypersetup{colorlinks,linkcolor={DarkBlue},citecolor={DarkBlue},urlcolor={DarkBlue}}

\usepackage{microtype}
\usepackage[caption=false]{subfig}
\usepackage{dsfont}
\usepackage{braket}
\usepackage{siunitx}

\usepackage[backgroundcolor = LightSteelBlue]{todonotes}

\newcommand{\e}{\operatorname{e}}
\newcommand{\diag}{\operatorname{diag}}

\begin{document}

\title{Hybrid spatiotemporal architectures for universal linear optics}

\author{Daiqin Su}
\thanks{These two authors contributed equally}
\affiliation{Xanadu, Toronto, Canada}
\author{Ish Dhand}
\thanks{These two authors contributed equally}
\affiliation{Xanadu, Toronto, Canada}
\affiliation{Institut f{\"u}r Theoretische Physik and Center for Integrated Quantum Science and Technology (IQST),\\
Albert-Einstein-Allee 11, Universit{\"a}t Ulm, 89069 Ulm, Germany}
\author{Lukas G. Helt}
\affiliation{Xanadu, Toronto, Canada}
\author{Zachary Vernon}
\affiliation{Xanadu, Toronto, Canada}
\author{Kamil Br\'adler}
\affiliation{Xanadu, Toronto, Canada}

\date{\today}

\begin{abstract}
We present two hybrid linear-optical architectures that simultaneously exploit spatial and temporal degrees of freedom of light to effect arbitrary discrete unitary transformations.
Our architectures combine the benefits of spatial implementations of linear optics, namely low loss and parallel operation, with those of temporal implementations, namely modest resource requirements and access to transformations of potentially unbounded size.
We arrive at our architectures by devising and employing decompositions of large discrete unitary transformations into smaller ones, decompositions we expect to have broad utility beyond spatio-temporal linear optics.
We show that hybrid architectures promise important advantages over both spatial-only and temporal-only architectures.
\end{abstract}

\maketitle

\section{Introduction}

Scalable linear optical interferometers are key to unlocking important quantum technologies such as quantum computation~\cite{Knill2001,Rudolph2017}, quantum metrology~\cite{Motes2015}, quantum simulations (including vibronic spectroscopy \cite{Huh2015}), boson sampling~\cite{Aaronson2013}, and Gaussian boson sampling~\cite{Hamilton2017}.
Obtaining a quantum advantage in these applications requires overcoming the challenging problem of scaling to a large number of photons and modes.
For instance, recent classical algorithms have raised the bar for a quantum advantage in boson sampling to requiring more than $30$--$50$ indistinguishable photons in thousands of modes~\cite{Neville2017,Clifford2018}.

Implementing a programmable linear-optical interferometer on thousands of modes is infeasible in bulk optics because of stability requirements.
An alternative is to integrate the components onto a monolithic photonic chip, which alleviates the stability issues~\cite{Carolan2015,Harris2016,Harris2017}.
However, scaling these chips to include a large number of modes is challenging as the dimensions of these chips are limited, and the optical elements, as well as their corresponding classical control elements, have a certain minimum-area footprint.
Indeed, whereas achieving a verifiable quantum advantage via boson sampling in integrated chips could require $10^{6}$--$10^{7}$ optical elements~\cite{Neville2017,Clifford2018}, current implementations are limited to tens of modes or hundreds of elements~\cite{[{See Table 3 in }]Flamini2019}.

Exploiting the potentially unbounded temporal degree of freedom provides an attractive alternative to using only the spatial modes of light on an integrated chip.
Existing architectures allow for implementing arbitrary discrete unitary transformations on the temporal modes of light in a single spatial mode~\cite{Motes2014,Motes2015a,takeda2017universal}.
Fully connected implementations of up to eight modes~\cite{He2017} and partially connected implementations of up to a million modes~\cite{Yoshikawa2016} have been demonstrated.
Although scaling to an arbitrarily large number of modes using only a limited number of optical elements is possible, in principle, using this architecture, two primary challenges remain.
First, this architecture requires using optical delay lines with lengths of the order of tens of kilometers~\footnote{This is for the \num{2500} modes required for verifiable quantum advantage with 50-photon boson sampling assuming current optical technology.}, which can induce significant loss and other imperfections as modes are switched in, propagated through, and switched out.
Second, the overall rate of performing a single run of the experiment decreases as the number of implemented modes is increased.

To date, only the two extremes of fully spatial or fully temporal interferometers have been explored, each with their individual advantages and shortcomings.
In this paper, we present two hybrid architectures that exploit both the spatial and temporal degrees of freedom of light to effect arbitrary discrete unitary transformations.
These hybrid architectures allow the exploration of the middle ground between spatial and temporal architectures and optimal trade-offs for specific implementations, enabling the construction of large interferometers with minimal experimental challenge.
Our architectures achieve this by combining the advantages of spatial implementations, i.e.,~lower switching and transmission losses and parallel operation, with those of temporal implementations, i.e.,~reduced number of optical elements and the possibility of implementing arbitrarily large transformations.

In devising the architectures, we detail two decompositions of $\text{SU}(N)$ transformations into products of $\text{U}(M)$ transformations.
Our hybrid architecture uses these decompositions to effect arbitrary $N\times N$ discrete unitary transformations on the temporal modes of light in $M$ spatial modes.
These spatial modes are acted upon by the obtained $\text{U}(M)$ transformations at different times, with optical delay lines connecting together different temporal modes within these spatial modes.

The rest of the paper is organized as follows. In Sec.~\ref{sec:decomposition}, we introduce two schemes to decompose $\text{SU}(N)$ transformations into
products of $\text{U}(M)$ transformations. Based on these decompositions, we propose hybrid spatiotemporal architectures to implement a large linear interferometer
in Sec.~\ref{sec:hybrid-architectures}. To demonstrate the advantages of the hybrid architectures, we compare the photon loss between the hybrid architectures and
the purely temporal architectures in Sec.~\ref{sec:photon-loss}. Finally, we conclude in Sec.~\ref{sec:conclusion}.

\section{Unitary transformation decompositions}\label{sec:decomposition}

Before detailing the decompositions and resulting architectures, we first introduce some architecture-independent definitions.
An $N$-mode linear-optical interferometer is characterized by a unitary operator ${U}$ that transforms $N$ bosonic annihilation operators $ a_1,  a_2, \dots,  a_{N}$ according to
\begin{align}
a_i \to  a^{\prime}_i = {U}^{\dag}  a_i {U} = \sum_{j=1}^{N} U_{ij}  a_j,
\end{align}
where $\bm{U} \equiv U_{ij}$ is an $N \times N$ unitary matrix.
If the global phase of the emitted light is inconsequential, we can assume that the matrix $\bm{U}$ is in $\text{SU}(N)$, the group of special unitary matrices.
Existing architectures for implementing a given $\text{SU}(N)$ matrix $\bm{U}$ via linear optics rely on first systematically decomposing $\bm{U}$ into $\text{U}(2)$ matrices, which are then identified as beamsplitter and phase transformations acting on different modes of light~\cite{Reck1994,Clements2016,Guise2018}.

\begin{figure*}
\subfloat[\label{Fig:Elimination}]{
\includegraphics[width=0.75\textwidth]{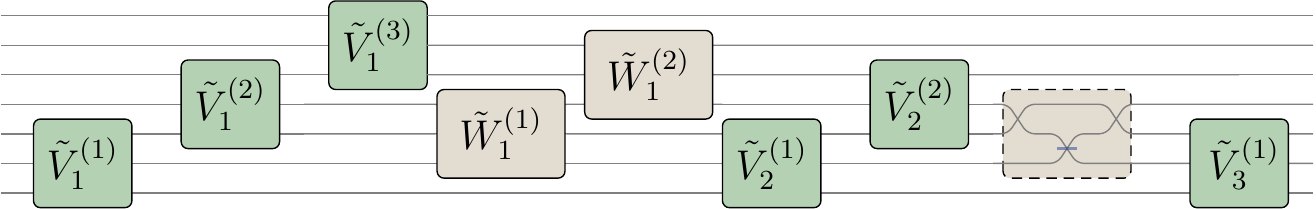}
}\\
\subfloat[\label{Fig:CSDbased}]{
\includegraphics[width = 0.9\textwidth]{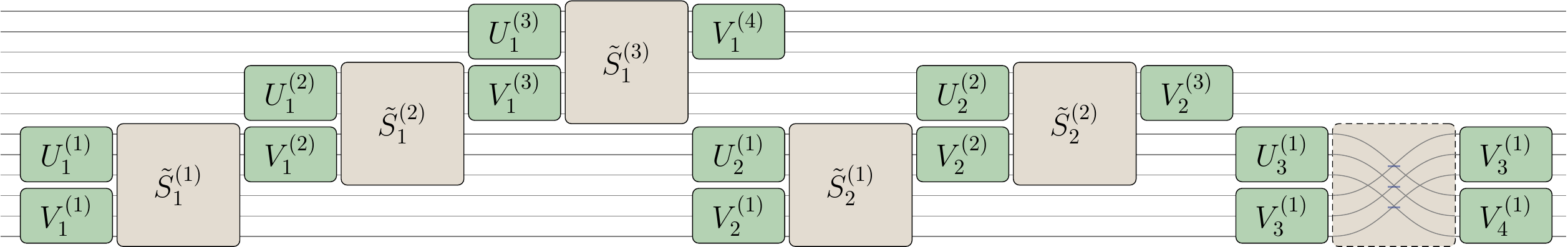}
}
\caption{
\textbf{Depiction of the two decompositions}.
(a) Elimination-based decomposition of an $N\times N$ unitary matrix into elementary matrices, including $M \times M$ universal unitary matrices (green) and specialized  $(2M-3) \times(2M-3)$ unitary matrices (brown) for $N = 7$ and $M = 3$.
(b) CS-based decomposition of an $N\times N$ unitary matrix into elementary matrices, including $M \times M$ universal unitary matrices (green) and specialized $2M \times 2M$ CS matrices (brown) for $N = 12$ and $M = 3$.
In each (a) and (b), the subscript labels the layer that each unitary block belongs to, and the superscript distinguishes different unitary matrices within each layer.
}
\label{Fig:Decompositions}
\end{figure*}

\subsection{Elimination-based decomposition}\label{sec:Elimination}

Here we present a decomposition of $\text{SU}(N)$ into smaller $\text{U}(M)$ and $\text{U}(2M-3)$ matrices.
Specifically, the procedure decomposes any given $\text{SU}(N)$ matrix into two types of elementary matrices: standard $\text{U}(M)$ matrices and specialized ``residual'' $(2M-3)\times(2M-3)$ unitary matrices, as illustrated in~\ref{Fig:Elimination}.
In this section we first illustrate the overall decomposition for the case of $N = 7$ and $M = 3$ before describing the general case of arbitrary $M$ and $N$.

For the purpose of elimination, we define an $N \times N$ unitary matrix $\bm{T}_{mn}(\theta,\phi)$ with $n>m$ following earlier work~\cite{Reck1994,Clements2016}.
$\bm{T}_{mn}(\theta,\phi)$ is obtained from the $N\times N$ identity matrix $\mathds{I}_{N}$ by changing the entries at the intersection of the $m$th and $n$th rows and columns to
\begin{align}
	\begin{pmatrix}
	\cos{\theta}\,\e^{-i\phi} & -\sin{\theta} \\
	\sin{\theta}\,\e^{-i\phi} & \cos{\theta}
	\end{pmatrix},
\end{align}
and leaving the other entries unchanged.
These matrices, with suitably chosen values of $\theta$ and $\phi$, can be multiplied with other matrices from the right in order to obtain new matrices in which specific entries are zero~\cite{Reck1994,Clements2016}.
Physically, each $\bm{T}_{mn}(\theta,\phi)$ can be realized using a tunable beam splitter and tunable phase shifter parameterized by $\theta$ and $\phi$ acting on modes labeled $m$ and $n$.
Henceforth, we drop the arguments~$(\theta,\phi)$ for simplicity.

We illustrate our procedure with the concrete example of decomposing an $\text{SU}(7)$ matrix into $\text{U}(3)$ matrices.
A general $\text{SU}(7)$ matrix ${\bm{U}}$ is represented by
\begin{align}\label{eq:SU7matrix}
\left(\begin{array}{c|c|c|c|c|c|c}
*& C_{(1,2)}^{8}& C_{(2,3)}^{7}& B_{(3,4)}^{5}& B_{(4,5)}^{4}& A_{(5,6)}^{2}& A_{(6,7)}^{1}\\
\hline
&*& C_{(2,3)}^{9}& E_{(2,4)}^{11}& B_{(4,5)}^{6}& D_{(4,6)}^{10}& A_{(6,7)}^{3}\\
\hline
&& *& G_{(3,4)}^{16}& G_{(4,5)}^{15}& F_{(5,6)}^{13}& F_{(6,7)}^{12}\\
\hline
&&&*& G_{(4,5)}^{17}& H_{(4,6)}^{18}& F_{(6,7)}^{14}\\
\hline
&&&&*& I_{(5,6)}^{20}& I_{(6,7)}^{19}\\
\hline
&&&&&*& I_{(6,7)}^{21}\\
\hline
&&&&&&*
\end{array}\right),
\end{align}
where the bottom off-diagonal part is not explicitly shown for simplicity and the elements are complex valued in general.
The matrix elements with subscripts $(m,n)$ above the diagonal are nulled systematically in the order of their superscripts using ${\bm T}_{mn}$ matrices.
The first step of the decomposition nulls the two rows as follows:
\begin{enumerate}
\item Multiply $\bm{U}$ by $(\bm{T}_{67} \bm{T}_{56} \bm{T}_{67})^{-1}$ to get ${\bm{U}}^{(1)}$ for
which the three entries labeled by ``$A$" in matrix \eqref{eq:SU7matrix} are zero.
\item Multiply ${\bm{U}}^{(1)}$ by $(\bm{T}_{45} \bm{T}_{34} \bm{T}_{45})^{-1}$ to
get ${\bm{U}}^{(2)}$ for which the three entries labeled by ``$B$" in matrix \eqref{eq:SU7matrix} are zero.
\item Multiply ${\bm{U}}^{(2)}$ by $(\bm{T}_{23} \bm{T}_{12} \bm{T}_{23})^{-1}$ to get
${\bm{U}}^{(3)}$ for which the three entries labeled by ``$C$" in matrix \eqref{eq:SU7matrix} are zero.
\item Multiply ${\bm{U}}^{(3)}$ by $\bm{T}_{46}^{-1}$ to get ${\bm{U}}^{(4)}$ for which the entry $U_{26}$ labeled by
``$D$" in matrix \eqref{eq:SU7matrix} is set to be zero.
 \item Multiply ${\bm{U}}^{(4)}$ by $\bm{T}_{24}^{-1}$ to get ${\bm{U}}^{(5)}$ for which the entry $U_{24}$ labeled by
``$E$" in matrix \eqref{eq:SU7matrix} is set to be zero.
\end{enumerate}

Note that the parameters of each $\bm{T}_{mn}$ above are chosen to null a specific matrix element and are, in general, not the same even if they have the same subscripts $m$ and $n$.
After these five smaller steps, which together comprise the first round of the decomposition, all entries in the first two rows of the top off-diagonal part are zero.
We continue the above procedure to set all entries in the top off-diagonal part to be zero, i.e., convert into lower triangular form.
The whole process can be described by
\begin{align}
\bm{U} \, (\bm{T}_{67} \bm{T}_{56} \bm{T}_{67})^{-1}
(\bm{T}_{45} \bm{T}_{34} \bm{T}_{45})^{-1}
(\bm{T}_{23} \bm{T}_{12} \bm{T}_{23})^{-1}
&\nonumber\\
\bm{T}_{46}^{-1}\bm{T}_{24}^{-1}
(\bm{T}_{67} \bm{T}_{56} \bm{T}_{67})^{-1}
(\bm{T}_{45} \bm{T}_{34} \bm{T}_{45})^{-1}
&\label{Eq:DecompositionU7}\\
\bm{T}_{46}^{-1}
(\bm{T}_{67} \bm{T}_{56} \bm{T}_{67})^{-1}
&= \bm{D},\nonumber
\end{align}
which gives $\bm{U}$ in terms of $\bm{T}_{mn}$ and diagonal matrix $\bm{D} = \text{diag}{(\e^{i \delta_1}, \e^{i \delta_2}, \cdots, \e^{i \delta_7})}$.
We know that $\bm{D}$ is diagonal because it is simultaneously unitary and in lower-triangular form.
The factors in the brackets of \ref{Eq:DecompositionU7} are combinations that can be grouped together into $\text{U}(3)$ matrices, denoted by $\tilde{V}$, acting on three adjacent rows and leaving the other rows unchanged.
The complete decomposition is presented in \ref{Fig:Elimination}, with the green boxes representing the $\text{U}(3)$ matrices.

The remaining unitary matrices $\bm{T}_{46}$ and $\bm{T}_{24}$ represent single beam-splitter transformations.
These transformations, unlike the remainder of the transformations, do not represent a coupling between adjacent modes.
Thus, $\bm{T}_{46}$ can be implemented by swapping modes with indices $4$ and $5$, which allows mode $4$ to be coupled to mode $6$, and swapping modes $4$ and $5$ back to their original order after the interaction.
Similar swaps are required for implementing the single beam splitter $\bm{T}_{24}$.
The swap operation can be viewed as a beam splitter with unit transmissivity.
We can combine two swaps and a single beam splitter to form a three-mode unitary, which we call a ``residual'' unitary, denoted by $\tilde{W}$, and is depicted within the dashed brown box in \ref{Fig:Elimination}.
We emphasize that the three-mode residual unitary is not universal because only one beam splitter is required to be tunable and the other two are fixed.

Altogether, the $\text{SU}(7)$ matrix can be decomposed into six $\text{U}(3)$ matrices $\tilde{V}$,
three residual matrices $\tilde{W}$, and seven  phases corresponding to the diagonal matrix $\bm{D}$.
As $\bm{U}$ is a special unitary matrix, one of these phases can be set equal to unity so only six additional phase shifters are required.


The above procedure can be generalized to decompose an arbitrary $\text{SU}(N)$ matrix for general $M$.
The first step, analogous to the five smaller steps described above, is to null the top $M-1$ rows using standard and residual matrices.
In particular, $k = (N-1)/(M-1)$ standard $U(M)$ matrices are used to null the $k$ triangular units of $M-1$ rows and columns each.
These units are analogous to the two-row two-column units $A, B, C$ in \ref{eq:SU7matrix}.
The remaining elements are nulled using the residual unitary $\tilde W$, the determination of which is described below.
Following this, we are left with a matrix that is diagonal in the top $M-1$ rows and unitary in the remaining rows and columns.
The next $M-1$ rows are nulled in the next step.
The procedure is completed after $k$ such steps, with each nulling $M-1$ rows.
Therefore, any given $\bm{U} \in \text{SU}(N)$ can be decomposed into $k(k+1)/2$ universal $\tilde V\in\text{U}(M)$ matrices and $k(k-1)/2$ residual matrices $\tilde W$ with size $(2M-3)\times(2M-3)$.

We illustrate the determination of general $\tilde W$ matrices with an example of decomposing a U($13$) unitary matrix into U($5$) unitary matrices.
After the application of the $\text{U}(5)$ unitary matrices, a top-right submatrix of an SU($13$) matrix is in the form
\begin{align}\label{eq:SuppResidueV}
\left(
\begin{array}{c|c|c|c|c|c|c|c}
0 & 0 & 0 & 0 & 0 & 0 & 0 & 0 \\
\hline
S_{(2,6)}^{7} & 0 & 0 & 0 & R_{(6,10)}^{1} & 0 & 0 & 0 \\
\hline
S_{(3,6)}^{9} & S_{(6,7)}^{8} & 0 & 0 & R_{(7,10)}^{3} & R_{(10,11)}^{2} & 0 & 0 \\
\hline
S_{(4,6)}^{12} & S_{(6,7)}^{11} & S_{(7,8)}^{10} & 0 & R_{(8,10)}^{6} & R_{(10,11)}^{5} & R_{(11,12)}^{4} & 0
\end{array}
\right).
\end{align}
A single residual unitary leads to the elimination of six nonzero entries $R_{(6,10)}^{1}$, $R_{(10,11)}^{2}$, $R_{(7,10)}^{3}$, $R_{(11,12)}^{4}$, $R_{(10,11)}^{5}$ and $R_{(8,10)}^{6}$, where the superscripts represent the order of elimination.
This elimination can be performed by sequentially applying $\boldsymbol{T}_{6, 10}^{-1}$, $\boldsymbol{T}_{10, 11}^{-1}$, $\boldsymbol{T}_{7, 10}^{-1}$, $\boldsymbol{T}_{11, 12}^{-1}$, $\boldsymbol{T}_{10, 11}^{-1}$, and $\boldsymbol{T}_{8, 10}^{-1}$.
These six beam-splitter transformations are combined together to form the specialized nonuniversal unitary $\tilde W$.

Note that similar to above, $\boldsymbol{T}_{6, 10}^{-1}$, $\boldsymbol{T}_{7, 10}^{-1}$, and $\boldsymbol{T}_{8, 10}^{-1}$ involve nonadjacent interactions. Thus, to physically implement $\boldsymbol{T}_{6, 10}^{-1}$, we need to shift the sixth port to the ninth port, which requires three swaps.
To physically implement $\boldsymbol{T}_{7, 10}^{-1}$ we also need three swaps since the seventh port has been shifted to the sixth port when implementing $\boldsymbol{T}_{6, 10}^{-1}$.
The same number of swaps is also required to implement $\boldsymbol{T}_{8, 10}^{-1}$.
After all entries are eliminated, another three swaps are needed to arrange all modes back to the original order.
Therefore, the total number of swaps is $4 \times 3 = 12$.

\begin{figure}[htbp]
\includegraphics[width=\columnwidth]{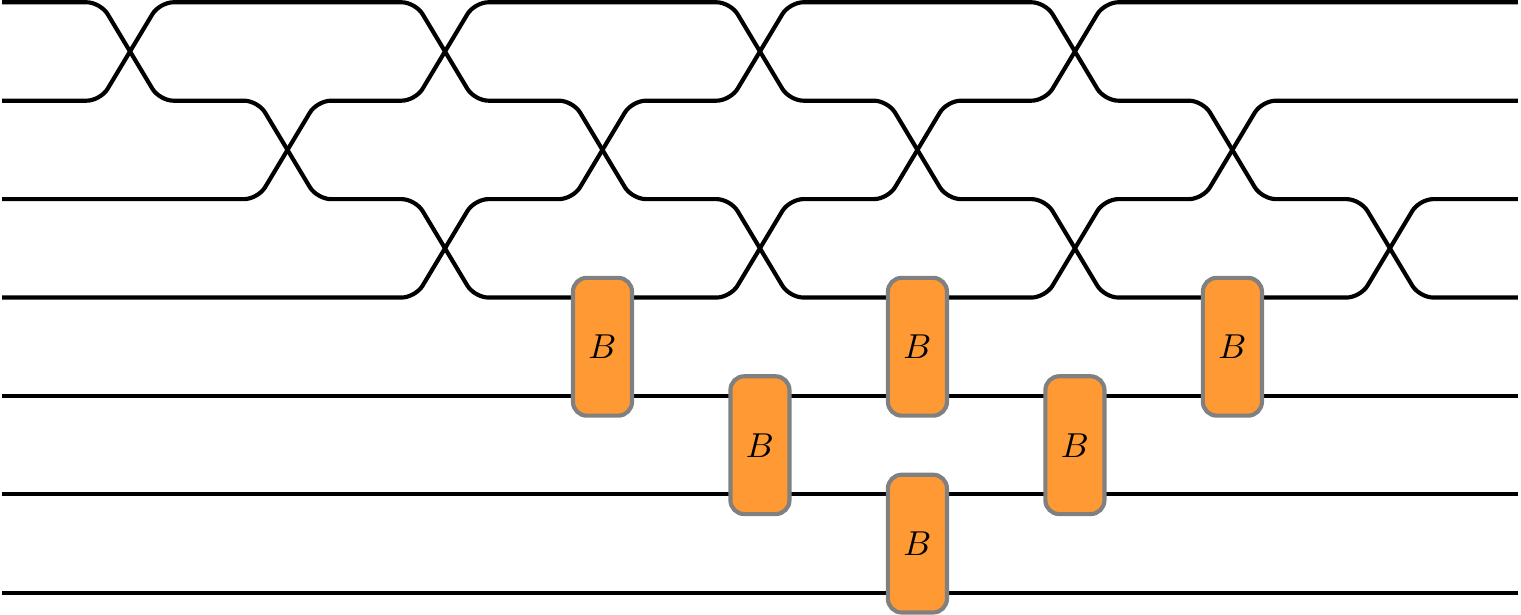}\caption{ A residual unitary $\tilde{W}$ for $M=5$. It consists of six tunable beam splitters (orange B-boxes) and 12 swaps.}
\label{fig:residual-UR-5}
\end{figure}

Figure~\ref{fig:residual-UR-5} shows the circuit that implements the residual unitary $\tilde W$ for $M=5$.
The above procedure to construct a residual unitary can be generalized to an arbitrary integer $M$ leading to a decomposition that requires $(M-1)(M-2)/2$ beam splitters and $(M-1)(M-2)$ swaps.

\subsection{Cosine-sine-based decomposition}

The second decomposition of $\text{SU}(N)$ into $\text{U}(M)$ that we consider is the decomposition presented in Ref.~\cite{Dhand2015}.
Although we use the same decomposition, our architecture presented differs from that of~\cite{Dhand2015}, which relies on implementing the obtained $\text{U}(M)$ matrices in the internal degrees of freedom of light.
In contrast, our motivation (as described in the next section) is to implement the $\text{U}(M)$ transformations in the spatial modes and reuse a single spatial interferometer over multiple passes using the temporal degree of freedom.

Here we recall the decomposition in more detail.
The decomposition is based on the cosine-sine decomposition (CSD), which factorizes any arbitrary $(m+n) \times (m+n)$ unitary matrix $U_{m+n}$ into unitary matrices $\tilde{L}_{m+n}$, $\tilde{S}_{2m}$, and $\tilde{R}_{m+n}$ according to~\cite{Stewart1977,Stewart1982,Sutton2009}
\begin{flalign}
U_{m+n}=& \tilde{L}_{m+n} \left(\tilde{S}_{2m}\oplus \mathds{I}_{n-m}\right)\tilde{R}_{m+n},\nonumber\\
\tilde{L}_{m+n} =&
\left(\begin{array}{c|c}
L_{m}& {0} \\
\hline
0 & L_{n}'
\end{array}\right),~
\tilde{R}_{m+n} =
\left(\begin{array}{c|c}
R^{\dagger}_{m}& {0} \\
\hline
0 & R^{\prime\dagger}_{n}
\end{array} \right),
\end{flalign}
where $\tilde{S}_{2m}$ is a cosine-sine (CS) matrix of the form
\begin{equation}
\begin{pmatrix}
\diag{(\cos \theta_{1}, \dots, \cos \theta_{m})} & \diag{(\sin \theta_{1}, \dots, \sin \theta_{m})}\\
-\diag{(\sin \theta_{1}, \dots, \sin \theta_{m})}  & \diag{(\cos \theta_{1}, \dots, \cos \theta_{m})}
\end{pmatrix}.
\label{Eq:CSMatrix}
\end{equation}
Note that the matrix subscripts give the dimensions of the matrices.
The matrices $L_{m}, L^{\prime}_{n}, R_{m}, R^{\prime}_{n}$ and the angles $\Theta = \{\theta_{1},\theta_{2},\dots,\theta_{m}\}$ can be determined using stable numerical methods~\cite{Dhand2015}. The CSD can be applied repeatedly to decompose an $N \times N$ unitary into smaller $M \times M$ universal unitary matrices and specialized (i.e., non universal) $2M \times 2M$ CS matrices, which are collectively referred to as elementary matrices~\cite{Dhand2015}.
The decomposition into elementary matrices, depicted in~\ref{Fig:CSDbased}, proceeds as follows.

Taking $N = \ell M$ for integer $\ell$, the decomposition is an iterative process comprising $\ell -1$ iterations.
In the first iteration, the full $N \times N$ matrix is decomposed via $\ell-1$ applications of the CSD into a single $(N-\ell)$-dimensional unitary matrix along with a ``layer'' of elementary matrices comprising $2\ell-1$ $M \times M$ unitary matrices ($U_{i}^{(j)}, V_{i}^{(j)}$ in~\ref{Fig:CSDbased}) and $\ell-1$ CS matrices ($\tilde{S}_{i}^{(j)}$ in~\ref{Fig:CSDbased}).
The first layer is depicted in \ref{Fig:CSDbased} by the boxes with subscript $1$.
In general, the $(i+1)$th iteration uses the CSD to decompose the $(N-i\ell)$-dimensional unitary matrix into a layer of elementary matrices and a smaller $(N-\left(i+1\right)\ell)$dimensional unitary matrix, which is then decomposed in subsequent iterations.
Eventually, the full unitary is decomposed into $\ell$ layers of elementary matrices, of which the last layer comprises a single ${M\times M}$ universal unitary matrix.

\section{Hybrid spatio-temporal architectures}\label{sec:hybrid-architectures}

Based on these two decompositions, we present corresponding architectures for implementing an arbitrary $\text{SU}(N)$ matrix on the combined temporal and spatial modes of light.
The key insight behind our hybrid architectures is that the action of one layer of interferometers on spatial modes of light can be replaced by the action of a single tunable interferometer, with suitable delay lines, on the spatial and temporal modes of light.

\subsection{Elimination-based hybrid architecture}
First, we consider the elimination-based decomposition, which returns $k = (N-1)/(M-1)$ ``layers'' of $M$-mode universal matrices $\tilde{V}$ and $k-1$ layers of ($2M-3$)-mode residual unitary matrices $\tilde{W}$.
Different layers are labeled by different subscripts in~\ref{Fig:Elimination}.

\begin{figure*}[htbp]
\includegraphics[width=0.9\textwidth]{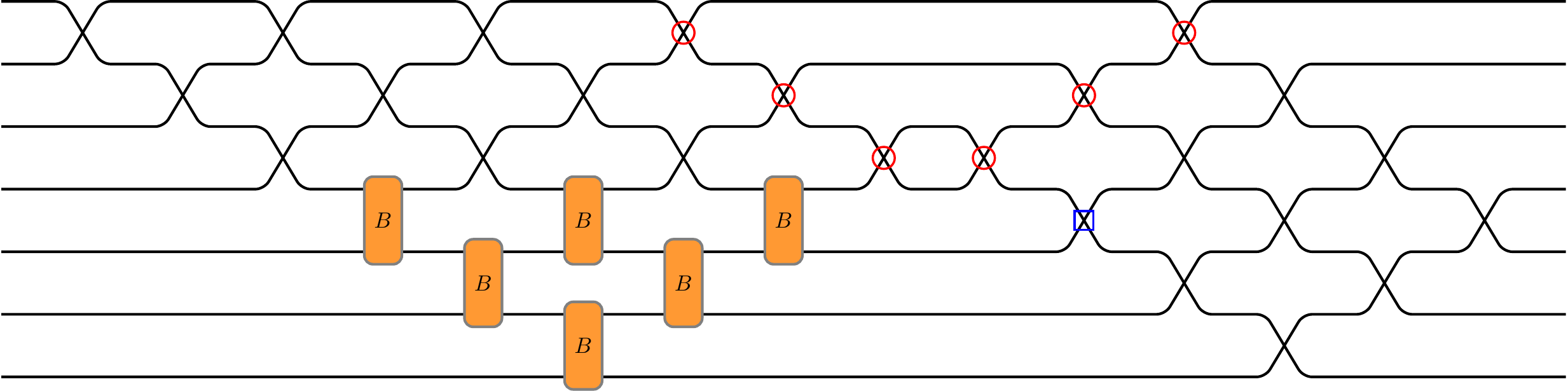}
\caption{ A residual unitary $W$ for $M=5$. To construct $W$ from $\tilde{W}$, another 12 swaps are needed. However, six swaps (red circles) cancel with each other and one swap (blue rectangle) can be absorbed into a tunable beam splitter.}
\label{fig:residual-UR-5-full}
\end{figure*}

To implement the hybrid architecture, one needs to shift the bottom $M-1$ rows of the $\tilde{V}$ and $\tilde W$ matrices to the top, and we denote the resultant matrices by $V$~and $W$.
A straightforward implementation of the resulting $V$ and $W$ matrices would require $(M-1)^{2}$ additional swap gates as compared to implementing $\tilde{V}$ and $\tilde W$.
Specifically, changing $\tilde{V}$ to $V$ straightforwardly requires implementing $(M-1)$ additional swaps.
However, these swap gates can be absorbed into $\tilde{V}$ to form another universal interferometer $V$ so no additional swaps are required.
From the residual unitary $\tilde W$ to $W$, we have to swap the bottom $(M-1)$ rows in place with the top $(M-1)$ rows.
Implementing this straightforwardly would require $(M-1)(M-2)$ swaps.
However, some of the swaps cancel with each other and the total number of swaps in $W$ is reduced.
In particular, at least $2M-3$ swaps can be canceled or absorbed into the beam splitters.
Figure~\ref{fig:residual-UR-5-full} shows the circuit that implements the residual unitary $W$ for $M=5$.

\begin{figure}
\subfloat[\label{Fig:EliminationSingle}]{\includegraphics[width = 0.8\columnwidth]{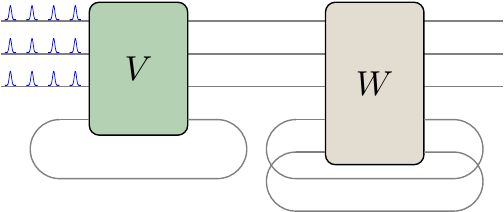}}\\
\subfloat[\label{Fig:CSDTwo}]{\includegraphics[width = 0.5\columnwidth]{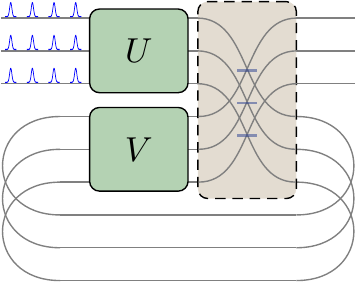}\label{Fig:S}}\hspace*{0.01\columnwidth}
\subfloat[\label{Fig:CSDSingle}]{\includegraphics[width = 0.5\columnwidth]{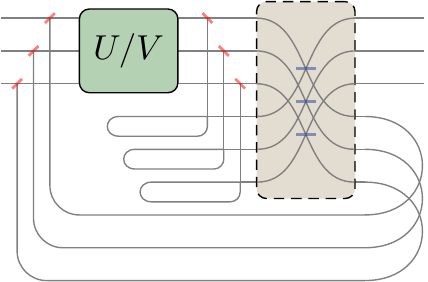}\label{Fig:Reuse}}
\caption{
(a)
Elimination-based decomposition realization of an $N\times N$ unitary into elementary interferometers comprising $M$-mode universal interferometers and $(2M-3)$-mode specialized interferometers for $M = 4$.
(b) CS-based realization of a single layer of elementary matrices on $N/M$ pulses in $M$ spatial modes for $M = 3$ using two universal interferometers and one nonuniversal interferometer comprising $M$ beam splitters and $M(M-1)$ swaps.
(c) Realization of both $U$ and $V$ interferometers using a single interferometer.
The blue (horizontal) dashes represent tunable beam splitters and the red (angled) dashes represent switches.
The delay lines at the bottom of (a) and (b) are assumed to effect a time delay $\tau$ equal to the pulse separation.
The two kinds of delays in (c) correspond to equal $\tau/2$ time delays before and after the $S_{2M}$ interferometer.
}
\label{Fig:Block}
\end{figure}

The hybrid architecture relies on implementing layers of matrices using individual tunable interferometers.
Consider $V$ a single tunable $M$-mode universal spatial interferometer that has $M-1$ free input and output ports and one output port connected to one input port with a delay line of length equal to the separation $\tau$ between subsequent temporal modes.
As detailed below, the single interferometer $V$ in \ref{Fig:EliminationSingle} enacts one layer of matrices $\tilde{V}_{j}^{i}$ for $i = 1,2,\dots,k$ in \ref{Fig:Elimination}.
Implementing $V$ requires $M(M-1)/2$ tunable beamsplitters~\cite{Reck1994,Clements2016}.

Similarly, the residual $W$ matrices are implemented using a specialized $(2M-3)$-dimensional spatial-mode interferometer and $M-2$ delay lines.
As already detailed in Sec.~\ref{sec:Elimination}, realizing these residual unitaries requires $(M-1)(M-2)/2$ tunable beamsplitters.

The sequence of operations for effecting a single layer of $V$ matrices is as follows.
Initially, a single temporal mode (pulse) impinges on $V$ at the first port and the $V$ interferometer is set such that the pulse moves into the delay line.
When this pulse is guided to the $M$th input of $V$, another $M-1$ pulses impinge simultaneously via the first $M-1$ inputs of $V$.
Then the first $M$-mode unitary $V_{1}^{(1)}$ is implemented.
After this action, the first $M-1$ output pulses from $V$ move on to the next layer, while the last output pulse moves into the delay loop and will couple with another $M-1$ pulses that arrive after an interval $\tau$ on unitary $V_{1}^{(2)}$.
This process continues until all $V_{1}^{(j)}$ unitary cells in the first layer are implemented.
A similar sequence of operations effects the $W_{1}$ layer.

The full $N \times N$ unitary matrix is a composition of the action of $k$ layers of $V$ matrices and $k-1$ layers of $W$ matrices.
Multiple layers of unitary matrices can be realized either by reusing these two interferometers using a dual-loop architecture along the lines of Refs.~\cite{Motes2014,Motes2015a} or by chaining together $k$ pairs of such interferometers in series~\cite{qi2018linear}.
In the dual-loop architecture, only a single block ( as in~\ref{Fig:EliminationSingle}) of two interferometers is required.
Also, a total of $M-1$ optical delay lines are used to feed the light emitted from the block back into the input of the block.
These delay lines, which implement time delays $\ge k\tau$, are attached to the output ports via switches that can guide some of the pulses into the delay lines while other pulses are transmitted onwards.
Each action of the two interferometers (\ref{Fig:EliminationSingle}) effects a single layer.
Thus, implementing a total of $k$ layers requires $k$ passes of the pulses through the $V$ and $W$ interferometers and $k -1$ passes through the delay lines.


\subsection{Cosine-sine-based hybrid architecture}
The second, CS-based architecture effects $\text{SU}(N)$ transformations on $M$ spatial and $\ell = N/M$ temporal modes of light.
The scheme employs tunable universal interferometers, each acting on $M$ spatial modes and nonuniversal $2M$-mode interferometers, each requiring only $M$ beam splitters.
A single layer of unitary blocks can be implemented on spatial and temporal modes using three optical elements (\ref{Fig:CSDTwo}).
The matrices $U$ and $V$ are universal tunable interferometers.
Finally, we replace $\Theta$ parametrizing each of the CS matrices obtained from \ref{Eq:CSMatrix} by their respective complements $\Theta^{\prime} = \{\pi/2-\theta_{1},\pi/2-\theta_{2},\dots,\pi/2-\theta_{M}\}$, thus replacing the obtained $\tilde{S}(\Theta)$ matrices by $S(\Theta^{\prime})$. This replacement eliminates the need for $M^{2}$ swap gates that would otherwise be required in the CS-based hybrid spatial-temporal architecture.
The $S$ matrix  can be realized using $M$ beamsplitters.

The sequence of operations for this architecture is as follows.
The first $M$ pulses, one in each of the $M$ spatial modes, arrive simultaneously at $U$.
On the first set of pulses, $U$ implements an identity transformation, letting these pulses pass unchanged.
Also only for the first set of pulses, $S$ redirects the pulses into the $M$ delay lines by tuning all the beamsplitters to unit transmissivity $T = \sin{\pi/2} = 1$.
Because of the delay lines, the next $M$ pulses arrive at $U$ at the same time that these cycling pulses arrive at $V$.
Now these two interferometers enact the first two unitary transformations of the decomposition procedure, i.e., the blocks $U_{1}^{(1)}$ and $V_{1}^{(1)}$ in Fig.~\ref{Fig:CSDbased}.
Together, these $2M$ pulses are acted upon by $S$.
Note that $M$ of these $2M$ pulses leave the interferometer and $M$ pulses enter the delay lines to arrive at the interferometers synchronously with the next set of $M$ pulses.
In the next round, the two universal and one nonuniversal interferometers are tuned to their next values, i.e., those corresponding to the next superscript in the first layer of Fig.~\ref{Fig:CSDbased}.
This process is repeated $\ell-1$ times, until the complete first layer is implemented.
As in the case of the elimination-based decomposition, the full unitary is implemented by concatenating multiple layers.
This is performed by chaining together a sequence of $\ell$ spatial interferometers one after the other, or using an appropriate dual-loop architecture~\cite{Motes2014,Motes2015a}.

Furthermore, the number of optical elements required to implement a single layer can be reduced.
In particular, two universal interferometers can be implemented by a single interferometer if $2M$ additional switches are available.
In this implementation (\ref{Fig:Reuse}), a single tunable interferometer plays the role of both $U_{M}$ and $V_{M}$ by switching between these two operating states after time $\tau/2$.
Two sets of $M$ optical lines implementing time delays of $\tau/2$ are used: one from the universal interferometer to the CS interferometer and another from the CS interferometer to the universal interferometer.

\section{Comparison of photon loss}\label{sec:photon-loss}

To show the potential advantages of the hybrid spatiotemporal architectures over other architectures,
we make a comparison of photon loss between a general hybrid architecture and a general fully temporal architecture. To characterize the loss of an architecture, we define the overall transmission as the transmission coefficient when the architecture implements an identity unitary transformation.
For the temporal architecture, the main sources of photon loss are the propagation loss in the inner loop and outer loop, loss in the tunable beam splitter, and the switching and coupling loss to the outer loop. We use the transmission coefficients $\eta_{\rm i}, \eta_{\rm o}, \eta_{\text{BS}}$, and $\eta_{\text{sc}}$ to represent these losses in the temporal architecture, respectively. Among these losses, the switching and coupling loss is dominant. To implement a layer of beam splitters, each pulse has to pass through the inner loop and the outer loop once, and the beam splitter and the switch twice. Therefore, the overall transmission coefficient when implementing an $N$-mode interferometer is
\begin{eqnarray}
\eta_{\rm{temporal}} = \big(\eta_{\rm i} \, \eta_{\rm o} \, \eta_{\text{BS}}^2 \, \eta_{\text{sc}}^2 \big)^{N-1}\eta_{\rm o}^{-1},
\end{eqnarray}
where we have assumed that each mode has identical loss.

For the spatial-temporal hybrid architecture, the main sources of photon loss are similar. If the block unitary is integrated on chip, then the dominant loss would come from the coupling in and out of the chip. We use the transmission coefficients $\tilde{\eta}_{\rm i}$, $\tilde{\eta}_{\rm o}$, $\tilde{\eta}_{\text{BS}}$, and $\tilde{\eta}_{\text{c}}$ to represent, respectively, each of these losses in the hybrid architecture. To implement a layer of block unitaries, each pulse has to pass through the inner loop and the outer loop once, the coupler twice, and the beam splitter $2M$ times.
Therefore, the overall transmission coefficient when implementing an $N$-mode interferometer is
\begin{eqnarray}
\eta_{\rm{hybrid}} = \big(\tilde{\eta}_{\rm i} \, \tilde{\eta}_{\rm o} \, \tilde{\eta}_{\text{BS}}^{2M} \, \tilde{\eta}_{\text{c}}^2 \big)^{k-1}.
\end{eqnarray}
where $k = (N-1)/(M-1)$ is the number of layers of block unitaries.

\begin{figure}
\includegraphics[width=0.95\columnwidth]{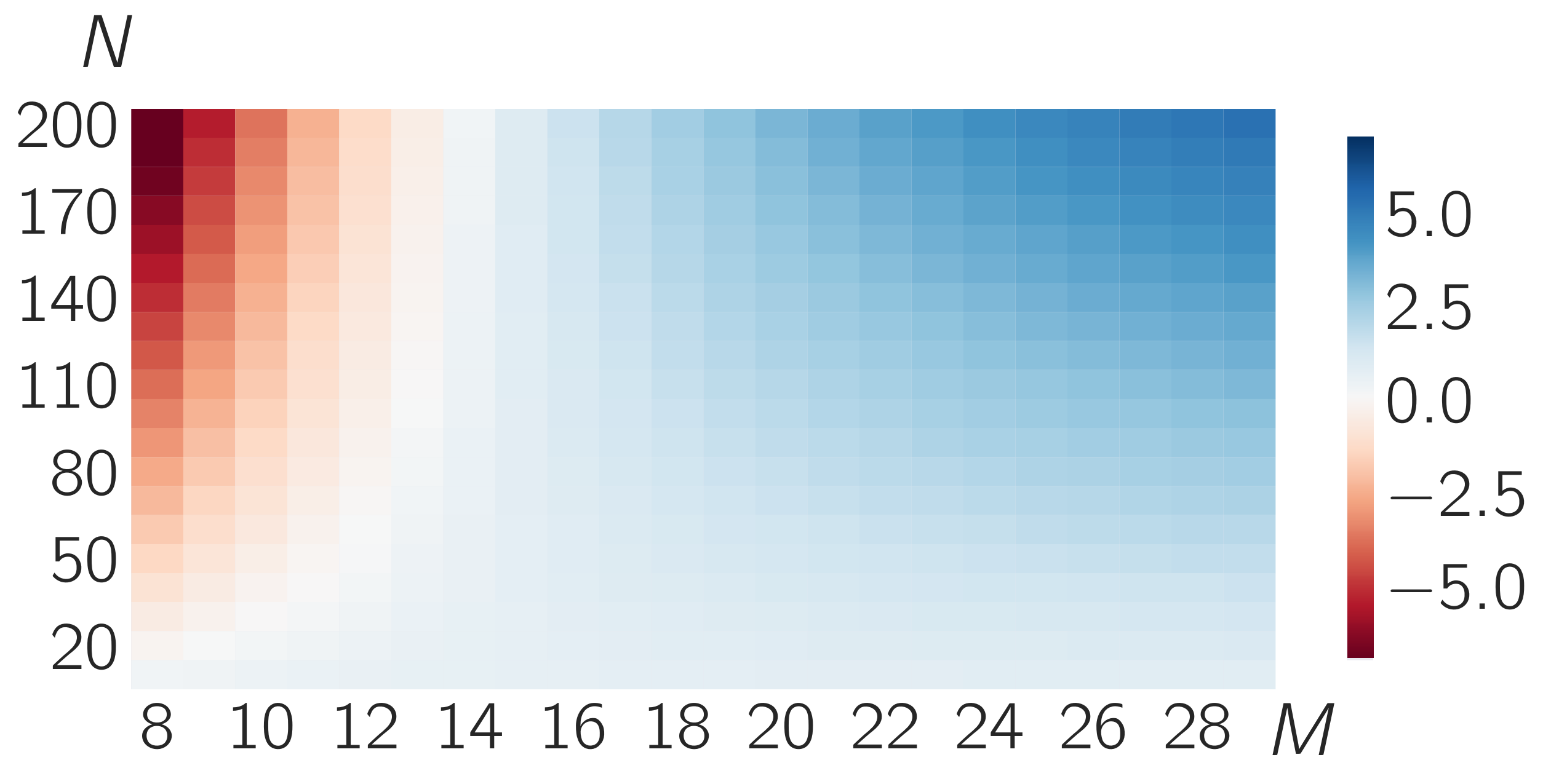}
\caption{
\textbf{Effect of loss in hybrid and temporal architectures}.
Plot of $\log_{10}(\eta_{\text{hybrid}}/\eta_{\text{temporal}})$ in the hybrid and temporal architectures for given $N$ and different choices of $M$, where $\eta$ is the transmissivity if $U = \mathds{I}$.
We assume realistic values of $50\%$ coupling efficiency to chip in the hybrid architecture and $95\%$ efficiency in switching and coupling to fiber in the temporal architecture.
}
\label{fig:ratio-overall-transmission}
\end{figure}

To make a comparison, we have to specify numeric values of the transmission coefficients, values that depend crucially on the platforms the architectures are implemented on. In general, we can assume $\tilde{\eta}_{\rm i} \approx \eta_{\rm i}$ and $\tilde{\eta}_{\text{BS}} \approx \eta_{\text{BS}}$. Since the outer loop of the temporal architecture is $k$ times longer than that of the hybrid architecture, we have ${\eta}_{\rm o} = \tilde{\eta}_{\rm o}^k$. The loss in the coupling in and out of the chip is usually higher than the loss in the coupling in and out of the outer loop (likely implemented in fiber), namely, $\tilde{\eta}_{\text{c}}<\eta_{\text{sc}}$. However, it is still possible that the overall loss of the hybrid architecture is lower than the temporal architecture. From the above assumptions, the ratio between $\eta_{\rm{hybrid}}$ and $\eta_{\rm{temporal}}$ is
\begin{eqnarray}\label{eq:supp:LossRatio}
\frac{\eta_{\rm{hybrid}}}{\eta_{\rm{temporal}}} &\approx& \eta_{\rm i}^{k-N} \, \eta_{\rm o}^{3-N} \eta_{\text{sc}}^{2(1-N)} \tilde{\eta}_{\text{c}}^{2(k-1)}
\approx \eta_{\text{sc}}^{2(1-N)} \tilde{\eta}_{\text{c}}^{2(k-1)}, \nonumber\\
\end{eqnarray}
where in the last step we assume that $\eta_i$ and $\eta_o$ are very close to one. From Eq.~\eqref{eq:supp:LossRatio}, it can be estimated that $\eta_{\rm{hybrid}}/\eta_{\rm{temporal}} \ge 1$ when $\tilde{\eta}_{\text{c}} \ge {\eta}_{\text{sc}}^M$. This means the overall loss of the hybrid architecture can be lower than that of the temporal architecture when $\tilde{\eta}_{\text{c}}$ is sufficiently large. As a concrete example, we compare $\eta_{\rm{hybrid}}$ and $\eta_{\rm{temporal}}$ by choosing representative values: $\eta_i = \eta_o = 0.9999, \eta_{sc} = 0.95$, $\eta_{\text{BS}}=0.96$, and $\tilde{\eta}_c = 0.5$, as shown in \ref{fig:ratio-overall-transmission}.

\section{Conclusion}\label{sec:conclusion}

To summarize, we have presented two architectures for the hybrid spatiotemporal implementation of a linear-optical interferometer based on decompositions of $\text{SU}(N)$ into products of $\text{U}(M)$.
These architectures also open the possibility of other hybrid architectures such as those involving frequency~\cite{Lu2018,reimer2019high} or orbital angular momenta~\cite{Garcia-Escartin2011} of light in addition to the spatial or temporal degrees of freedom.
Beyond optics, the architectures could be useful in realizing unitary transformations on other quantum systems such as ion traps and superconducting circuits if individual sites possess a multilevel structure~\cite{Shen2014,Peropadre2016,Goldstein2017}.
Especially for such implementations, an improvement of the current architecture to bring the $\text{U}(M)$ blocks into a rectangular form could be helpful.

Our hybrid spatiotemporal architectures fill the space between the two extremes of fully spatial and fully temporal architectures.
They maintain two advantages of temporal architectures, namely, a potentially unlimited number of realizable modes and a small number of required optical elements.
Our two decompositions lead to architectures that have $\mathcal{O}(N^{2}/M^{2})$  fewer optical elements than fully spatial realizations.
This reduction in elements comes at the experimental cost of having to stabilize $\mathcal{O}(M)$ delay lines and also with a concomitant $\mathcal{O}(N/M)$ increase in the time required to implement an $\text{SU}(N)$ transformation.

Furthermore, the hybrid architectures also allow for two advantages of fully spatial architectures which are not present in fully temporal architectures.
First, the parallel operation on $M$ modes of light leads to a factor $\mathcal{O}(M)$ speedup over fully temporal architectures.
Second, each of the pulses needs to cycle in the outer loop $\mathcal{O}(M)$ fewer times as compared to fully temporal architectures.
As a result, for large $M$, the hybrid architectures avoid losses associated with repeatedly cycling in the delay lines.
For the large number $N$ of modes required for demonstrating quantum advantage in boson sampling~\cite{Neville2017,Clifford2018}, a hybrid architecture promises many orders of magnitude improvement over an all-temporal architecture assuming realistic values of losses.
Thus, based on experimental capabilities and requirements, our architectures enable optimized implementations of $\text{SU}(N)$ unitary transformations.

\begin{acknowledgments}
\textbf{Acknowledgments:} The authors thank Christian Weedbrook, Casey Myers, Hubert de Guise, and especially Haoyu Qi for discussions and insightful comments.
I. D. acknowledges support from the Alexander von Humboldt Foundation via the Humboldt Research Fellowship for Postdoctoral Researchers and the Bundesministerium f\"ur Bildung
und Forschung project Q.Link.X.
\end{acknowledgments}

\bibliography{hybrid-st-decomposition.bbl}


\end{document}